# Efficient Indexing and Querying over Syntactically Annotated Trees

Pirooz Chubak
Department of Computing Science
University of Alberta
pchubak@ualberta.ca

Davood Rafiei
Department of Computing Science
University of Alberta
drafiei@ualberta.ca

## ABSTRACT

Natural language text corpora are often available as sets of syntactically parsed trees. A wide range of expressive tree queries are possible over such parsed trees that open a new avenue in searching over natural language text. They not only allow for querying roles and relationships within sentences, but also improve search effectiveness compared to flat keyword queries. One major drawback of current systems supporting querying over parsed text is the performance of evaluating queries over large data. In this paper we propose a novel indexing scheme over unique subtrees as index keys. We also propose a novel *root-split* coding scheme that stores subtree structural information only partially, thus reducing index size and improving querying performance. Our extensive set of experiments show that *root-split* coding reduces the index size of any interval coding which stores individual node numbers by a factor of 50% to 80%, depending on the sizes of subtrees indexed. Moreover, We show that our index using *root-split* coding, outperforms previous approaches by at least an order of magnitude in terms of the response time of queries.

## 1. INTRODUCTION

Natural language text is a prominent source of representing and communicating information and knowledge. It is often considered as an arbitrary collection or sequence of words by search systems. Such systems ignore the regularities and inherent structure that exist in finer granularity portions of text such as sentences. Tools such as part of speech taggers, syntactic parsers and semantic role labelers augment natural language text with tags and relationships that can be used to improve the effectiveness of searching over text significantly.

Various applications have shown interest in using the output of such text augmenting tools. A body of recent works [6, 18] focuses on improving the precision and recall of open information extraction systems by attending to the inherent structure underlying natural language text to train features for their extractions, and achieves up to %70 improvement on the F-measure of extractions compared to TextRunner [3]. Moreover, Question answering systems such as PowerAnswer [11] and MULDER [10] use syntactic parsing in several components of their system.

As an example of how such tools can improve the effectiveness of searching over natural language text, imagine one is interested in finding out the answer to the TREC-2004 question 'What kind of animal is agouti?' Using a keyword-based search engine. She can send a query such as 'agouti' and scan the returned results for the desired answer. Alternatively, if a corpus of syntactically parsed sentences exists, it can be matched against a parse of the query 'agouti is a'. As a result, sentences having the same structure and terms as the given parsed query will be returned in the result set. Figure 1 shows the parse tree of the query and the parse tree of a sample sentence containing the answer, parsed using the Stanford parser [9]. The matched subtree has been bolded in this figure for better visualization. As this figure shows, a few words separate the match, *rodent*, from the query terms. However, the parser manages to identify the correct syntactic relationship between the match, and the query terms.

### 1.1 Problem Statement

Despite great advantages of using parsing in terms of improving the search effectiveness, syntactic parsers are not in widespread use for querying natural language text. One reason that is sometimes cited is that parsing is a slow process. As an example, Stanford parser takes a few days to parse one million sentences on a single machine. However, this is not a major problem since parsing can be parallelized on a large number of machines, reducing its time cost significantly. Moreover, parsing is a one time task and the amortized cost of parsing a corpus of natural language text once and querying it many times, is not high.

The other and probably a stronger reason why syntactic parsing has not been widely used is that there is a lack of efficient storage mechanisms and access methods over large scale parsed text. Some of the major querying systems over syntactically parsed corpora such as TGrep2 [14] and CorpurSearch [13] require an in-memory scan of the entire corpus for answering any single query. As a result, they cannot scale to large corpora. LPath [4] and similar *relational node* approaches[1] that store structural information of individual nodes in a RDBMS, improve the performance of evaluating small or selective queries. However, they suffer from long posting lists over low selectivity node labels and require a large number of joins for evaluating large queries. To alleviate these problems *path* indexes (e.g. see [16, 15, 20]) and *frequent subtree (or subgraph)* indexes such as [23, 19] have been proposed over tree structured data and can be utilized for querying syntactically annotated trees. One major drawback of path indexes is that trees and graphs decomposed into paths lose structural information, and a unique tree or graph cannot be constructed from the decomposed paths. As a result, such systems either do not support exact query matching or require post validations to find the set of matching nodes. Indexes which use frequent sub-structures do not suffer from loss of structural information, but still require post validations for exact matchings, as non-frequent structures are not retained in the indexes. Moreover, such indexes often perform best when statistics over the frequent patterns in the queries are available. Finally, Williams et al. [17] index subgraphs of all sizes as index keys. They assume that input graphs are very small, which makes extraction of all subgraphs possible. Over parse trees which often have more than 100 nodes, extracting subtrees of all sizes takes prohibitive time. In this paper, our goal is to develop index structures and access methods that support exact query matching over parse trees and addresses the aforementioned problems. Our solutions exploit the inherent characteristics of natural language text to improve the performance of querying over syntactically parsed trees.

---
[1] often over XML documents

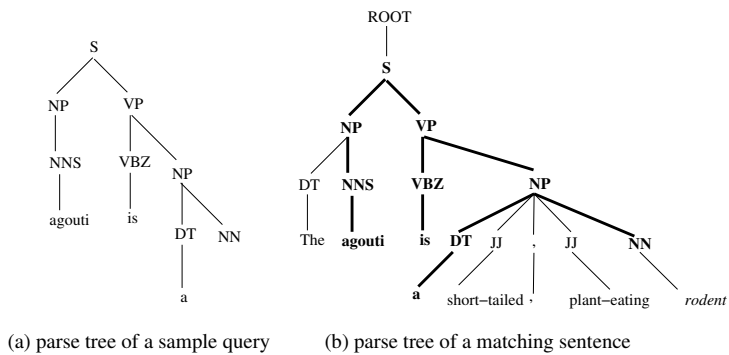

(a) parse tree of a sample query    (b) parse tree of a matching sentence

**Figure 1: A sample query tree and a parsed matching sentence. The matched subtree is bolded. Label nodes adhere to Penn treebank tags.**

## 1.2 Contributions and Paper Organization

The contributions of this paper are the following.

- We propose a Subtree Index (SI) over syntactically parsed corpora of natural language text. SI stores unique subtrees (up to a certain size) from the corpora as index keys. It also stores the structural information of each subtree in a set of posting lists which can be used to perform exact tree matchings. We show experimentally that SI can achieve a large query run-time speedup compared to the scenario where only structural information of nodes are stored (see [4] for an example). To the best of our knowledge, SI is the first work on indexing tree structured data that stores the set of unique subtrees as index keys.

- A novel root-split coding is proposed that concisely stores the structural information of subtrees in subtree index, making it possible to perform exact matching, while reducing the index size, index construction time and query response times, significantly.

- A query decomposition algorithm is proposed over baseline coding schemes that in our settings achieves optimality in terms of the number of joins required to evaluate queries. Also, our query decomposition algorithm over root-split coding achieves optimality among all root-split decompositions.

- We experimentally show that when SI stores subtrees larger than one node, it outperforms the node approach. We also show that our root-split coding outperforms our baseline coding schemes in terms of the query response times, is highly scalable and has a reasonably small index size and index construction time. Finally, we show that our index using root-split coding outperforms previous approaches supporting syntactically annotated tree queries by at least one order of magnitude.

The structure of the paper is as follows. In the next section, we study the literature around querying parse trees. We cover relevant work around query systems on syntactically parsed corpora, XML documents, trees and graphs. In Section 4 we provide the details of our subtree index and describe our baseline and root-split coding schemes. Section 5 theoretically studies the properties of our coding schemes and presents query decomposition algorithms. In Section 6 we evaluate the performance of our system using different sizes of subtrees and coding schemes and compare it with previous approaches. Finally, we conclude in Section 7 and provide a list of potential future avenues.

## 2. RELATED WORK

In the past couple of decades there has been an emergence of developing querying systems over syntactically annotated trees. Some of the major systems that operate on the output of syntactic parsers are TGrep2 [14] which is a grep for parse trees, CorpusSearch [13] and LPath [4]. TGrep2 and CorpusSearch, load the corpus in the main memory and scan the entire corpus to evaluate each query. Thus, their querying performance degrades over larger corpora and they cannot scale. LPath [4] uses an indexing approach to query syntactically annotated trees. It uses a relational database

1317

to store some structural information of nodes such as interval codes. LPath performs better than Tgrep2 and CorpusSearch, specifically over queries having labels with higher selectivities (See [4]). Our subtree index improves the performance of LPath by pre-materializing subtrees larger than single nodes, thus reducing the size of posting lists and the number of joins required to evaluate queries.

There has been extensive work on indexing and querying over trees and graphs. Shasha et al. [16, 15] proposed ATreeGrep, which facilitates approximate and exact matching over unordered trees. ATreeGrep stores all paths of the set of input trees into a suffix array. It also uses a hash index over all nodes and edges to pre-filter a set of candidate trees and to improve overall querying performance. Tree matching is done by decomposing the query tree into its root to leaf paths, and evaluating them against the suffix array. In contrast to our subtree index, ATreeGrep does not support distinct labels over different children of a node. It also does not support single node queries. Moreover, our subtree interval and root-split codings remove the need for post-validations. As a result, as confirmed by our experiments, our subtree index using root-split coding performs at least an order of magnitude faster than ATreeGrep. The work by Williams et al. [17] on subgraph isomorphism, stores the canonical forms of all subgraphs into an index. However, they assume that the input graphs are very small, making it possible to compute and store the exponentially many subgraphs of all sizes. Their approach cannot be used over parse trees as computing subtrees of all sizes is prohibitive in time. TreePi [23] uses frequent subtrees as elements of its index for subgraph isomorphism problem. TreePi prunes the search space of candidate graphs, and finds the set of matches using post validations. Compared to TreePi, our approach stores all subtrees up to a certain size and performs exact matching over the index. Moreover, our root-split and subtree interval codings do not require any post validations. As we discuss in Section 6, an adaptation of TreePi to indexing parse trees results in smaller index sizes, but a worse querying performance compared to our root-split coding.

There is also work on improving the performance of evaluating twig queries [5] by reducing the size of intermediate results while processing joins over streams of structural information stored in an index over XML documents. Multi-Predicate MerGe JoiN (MPMGJN) [22], StackTree [1] and TwigStack [5] are a few of the pioneer works on improving the performance of structural joins over twig queries (See [8] for a survey). Our implementation of subtree index in this paper, uses MPMGJN off-the-shelf for processing structural joins over subtrees. More efficient stack-based approaches can be directly applied over our root-split coding, but might require extra processing over our baseline approaches.

## 3. PRELIMINARIES

In this section we formally define our data and query models and the semantics of matchings.

DEFINITION 1. *A syntactically annotated tree $T$ is a tuple $T = \{V, E, \Sigma_V, m\}$, where $V$ is the set of nodes, $E \subseteq V \times V$ is a set of directed edges, $\Sigma_V$ is the set of node labels of $T$ and the function $m$ defines the mappings of $V \mapsto \Sigma_V$. $T$ is acyclic and is distinguished by its root, $r(T)$, where for every $v \in V$, there is a directed $r - v$ path in $T$. For any $(u, v) \in E$ edge, $u$ is called a parent node and $v$ is called a child node.* *For any node $v \in V$, it has a unique parent, except for $r(T)$ which has no parents.*

DEFINITION 2. *A tree query or simply a query $Q$ is a directed tree $Q = \{V, E, \Sigma_V, \Lambda_E, m\}$. $\Lambda_E$ is the set of edge types. Type of an edge could be any of* navigational axes. *Function $m$ defines mappings of $E \mapsto \Lambda_E$ as well as $V \mapsto \Sigma_V$.*

Navigational axes are a set of binary structural relationships between pairs of nodes in a query tree. Examples are parent-child axis, denoted by /, and ancestor-descendant axis, denoted by // (See [4] for a complete list of axes). Thus, $A//B/C$ indicates a query with root $A$ with a child $C$ and a descendant $B$. Note that queries are unordered, i.e. changing the order of $B$ and $C$ does not change the query semantics.

DEFINITION 3. *A query $Q = \{V, E, \Sigma_V, \Lambda_E, m\}$ matches a tree $T = \{V', E', \Sigma'_V, m'\}$ if there exists a mapping function $f : V(Q) \mapsto V'(T)$ that maps nodes of $Q$ to nodes of $T$, such that (1) $\forall v \in V, m(v) = m'(f(v))$ and (2) $\forall (u, v) \in E$, then $f(u)$ and $f(v))$ must be in the same relationship in $T$ as suggested by $m(u, v)$.*

To support large-scale query matching over tree structured data efficiently, a common practice is to assign a set of numbers to each node in the data tree and index the trees based on these numbers. Such numbers represent the *structural information* of nodes in the data tree. One such numbering scheme, commonly referred to as *(node) interval coding* [22], assigns to each data tree node a pair of *pre* and *post* numbers indicating the pre- and post-visit ranks of the node in a DFS traversal, respectively. In addition, a *level* number is assigned to handle parent-child queries. An inverted index is then constructed on *(treeId, pre, post, level)* values, sorted on the increasing order of *treeId* and *pre* values.

## 4. SUBTREE INDEX

In this section, we propose a novel subtree index and a few storage and querying techniques over this index. We present some structural properties of the index and an analytical study of its performance. Specifically, we study how interval coding can be extended to represent the structural information of subtrees; hence our subtree interval coding. We further introduce a novel root-split coding which leads to a more concise index compared to subtree interval coding and reduces the response time of queries as well as index construction time.

Given a set of syntactically annotated trees $S$ and a size parameter $mss$, consider the set of all unique subtrees of sizes $1, 2, \ldots, mss$ that can be extracted from trees in $S$, and associate to each subtree a posting list consisting of the IDs of trees in $S$ where the subtree appears. We want to organize the pairs of subtrees and posting lists in an index, referred to as *Subtree Index* (or *SI* for short), such that our queries can benefit from this structuring.

### 4.1 Subtree Indexes over Syntactically Parsed Trees

One drawback of subtree indexes is that their size could be prohibitively large. Two factors that affect the size of a subtree index are (1) the number of unique subtrees (index keys), and (2) the total number of extracted subtrees. This latter number gives an upper bound on the total number of postings in the index. Next, we study how the above two factors change over sample datasets of syntactically annotated text using a few experiments. We also discuss what properties of syntactically annotated trees make the size of a subtree index over them manageable.

*Number of Index Keys*

The number of index keys is equal to the number of unique subtrees extracted from the set of data trees. One nice property of syntactically annotated trees is that the number of index keys (and unique subtrees over them) grows almost linearly with the size of the input, for different values of $mss$. As a result, the body of the index does not grow dramatically as more data is being indexed. One reason for this is that similar structures are abundant throughout the corpus of parsed trees. This is based on the observation that there is only a finite and relatively small set of grammatical structures used in natural languages, and the number of such unique structures does not grow dramatically even considering differences in writing styles and parsing errors.

Figure 2 shows the number of unique subtrees as a function of the input size, for different values of $mss$, over collections of parse trees containing 1 to $10^6$ sentences from a news corpus. The figure shows approximately the same rate of growth in the number of keys, for different values of $mss$. Moreover, the number of index keys grows almost linearly with the size of the indexed data.

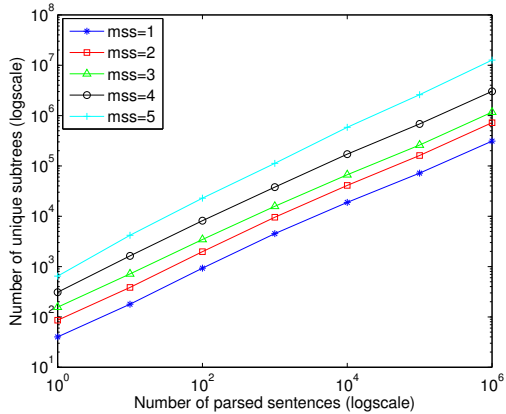

Figure 2: Number of index keys (unique subtrees) as a function of the input size in terms of the number of sentences

*Number of Extracted Subtrees*

For a tree of size $n$, the number of subtrees of size $m$ could range from $n - m + 1$ to $\binom{n-1}{m-1}$. The former belongs to the case where the tree is a unary branch of height $n$, and the latter demonstrates the case where the parse tree consists of a root with $n - 1$ leaf children. Note that the number of subtrees of sizes $1, \ldots, mss$ of a tree gives an upper bound on the number of postings stored for it in the index. Therefore, for large values of $mss$ and $n$, the number of postings stored in the index could be very large, resulting in a huge index. As we show later over syntactically annotated trees, the number of subtrees is in practice orders of magnitude smaller than the worst case scenario, making it possible to build SI for small values of $m$ (e.g. $1 \leq m \leq 5$).

To study how the number of extracted subtrees changes over syntactically annotated trees, we conducted an experiment on more than $50,000$ nodes from a (constituency) parsed corpus of news. Over each node, we extracted every possible subtree of sizes 2 to 5, and counted the number of such subtrees. Figure 3 depicts how the number of subtrees changes with the branching factor of the nodes, for this dataset. As this figure shows, nodes with higher branching factors lead to more subtrees, on the average.

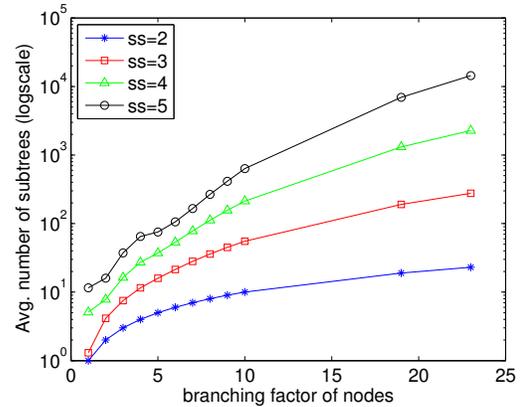

Figure 3: Average number of subtrees extracted in terms of the branching factor of roots of subtrees

Some of the important characteristics of syntactically annotated trees that distinguishes them from other tree structured data types are as follows. First, syntactically annotated trees have a small average branching factor. The average branching factor for internal nodes in the above dataset is 1.52. Thus, on average, each internal node has less than two children. Second, syntactically annotated trees rarely have nodes with large branching factors. In our above experiment, there exist only two nodes with branching factors larger than 10, while there could exist nodes with branching factors of a few hundreds or even larger over XML documents. This is due to two reasons. (a) Parse trees are relatively small trees and (b) high branching factor nodes in parse trees are due to highly repetitive structures which are rare in well-written natural language corpora. Finally, the above properties are fairly consistent across different corpora and syntactic parsers. Therefore, syntactically annotated trees are good candidates for subtree indexes as the number of index keys and posting list sizes are manageable for small values of $mss$.

## 4.2 SI Construction

A subtree index is parameterized by $mss$, the maximum subtree size. Given $mss$ and a set of parsed trees, SI extracts the set of all unique subtrees of sizes $1, \ldots, mss$, then flattens and encodes them as index keys. Figure 4 depicts an example of how unique subtrees of sizes 2 and 3 are extracted as keys of SI. The given input tree has 8 and 7 index keys of sizes 4 and 5, respectively. Note that the index keys are considered unordered. Thus, postings of `A(B)(C)` and `A(C)(B)` are stored under the same key entry in the index.

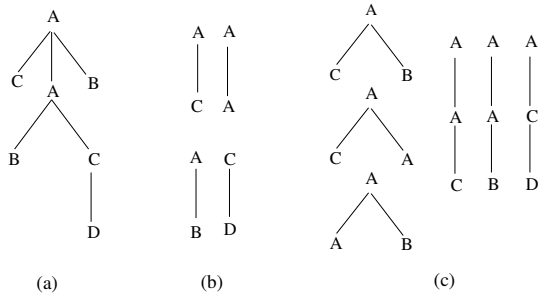

**Figure 4: Computing index keys, (a) input tree, index keys of sizes (b) 2, (c) 3**

Once index keys have been extracted, they need to be flattened, encoded and stored into the index. We traverse each subtree in pre-order and for each node capture its label and size. In this approach, a tree can be encoded by exactly $mss(\lceil \log_2(mss+1) \rceil + \lceil \log_2 |\Sigma_V| \rceil)$ bits. Recall that $\Sigma_V$ is the alphabet of node labels.

### 4.3 Query Matching over Subtree Indexes

Query matching over a subtree index has two phases, (1) the *query decomposition phase* in which the queries are decomposed into smaller subtrees, where each subtree size is at most *mss*, and the posting lists of subtrees are fetched from the index, and (2) the *join phase* in which these posting lists are joined to compute the final set of results.

DEFINITION 4. *For two trees $T$ and $T'$, we say that $T$ is a subtree of $T'$ and denote it by $T \preceq T'$ if and only if (1) $V(T) \subseteq V(T')$ and (2) $E(T) \subseteq E(T')$.*

DEFINITION 5. *A set $C = \{c_1, \ldots, c_k\}$ of trees is a* node-cover *of tree $T$, if and only if (1) for all $c_i \in C$, we have $c_i \preceq T$ and (2) for all $v \in V(T)$ there exists at least one $c_i \in C$ such that $v \in V(c_i)$.*

Intuitively, a *node-cover* of a tree $T$ is a set of subtrees of $T$ such that every node of $T$ appears on at least one of the subtrees of the node-cover.

DEFINITION 6. *A set $C = \{c_1, \ldots, c_k\}$ of trees is a* full-cover *of $T$, if and only if (1) $C$ is a node-cover of $T$, and (2) for all $e \in E(T)$ there exists at least one $c_i \in C$ such that $e \in E(c_i)$*

According to the above definition, a *full-cover* $C$ of tree $T$, covers both nodes and edges of $T$. Hereafter, we refer to both *full-covers* and *node-covers* simply as *covers*, when the meaning is clear from the context.

DEFINITION 7. *Given a query $Q$ and a size parameter mss, $C = \{c_1, \ldots, c_k\}$ is a valid cover of $Q$ with respect to mss if and only if $C$ is a cover of $Q$ and there does not exist a subtree $c_i \in C$ such that $|c_i| > mss$, for all $1 \le i \le k$.*

In the rest of the paper, we assume that all covers are valid, unless otherwise noted. Given a query, our goal in the *query decomposition phase* is to find a "good" cover. A "good" cover can be informally defined in terms of its closeness to a cover that results in the least query execution cost. A query could have a large number of covers, and the choice of which cover to pick can significantly affect query execution cost. In Section 5 we study a few properties of covers over proposed coding schemes that help us prune the search space of "good" covers. Next we discuss our coding schemes over subtree indexes and a brief overview of how query matching is performed over each coding scheme.

### 4.4 Coding Schemes

In this section we propose three coding schemes for encoding the structural information of subtrees stored as keys in a subtree index. The first two coding schemes are adaptations of current coding schemes over text or XML documents; they are mainly used as baseline methods. We also propose a novel root-split coding which stores the subtree structural information more concisely.

#### 4.4.1 Filter-Based Coding

The filter-based coding is a minimal coding scheme which does not store any structural information about the keys being indexed. Similar to any inverted index structure, the filter-based coding stores a sorted list of unique tree identifiers, *tid*s, of the trees that contain the indexed subtrees.

Query matching for the filter-based coding starts by finding a cover and fetching the respective posting lists of the subtrees in the cover. The join phase includes pairwise intersection of these lists to obtain the list of candidate *tid*s. Unlike the other two coding schemes, Query matching for the filter-based coding requires a third (usually costly) post-validation phase, called the *filtering phase*. In the filtering phase, the parse trees corresponding to candidate *tid*s are fetched and scanned to check if they match the query.

#### 4.4.2 Subtree Interval Coding

As discussed in Section 3, interval coding stores for each node a pair of *pre* and *post* values to handle *containment queries* [22] and a *level* value to answer parent-child axes queries. A subtree interval coding generalizes the node interval coding and stores for each indexed subtree the *(pre, post, level, order)* numbers of all of its nodes, as the structural information of the subtree.

The *order* value, the order of a node in a pre-order traversal, is stored to differentiate between instances of symmetric postings, which are stored under the same index key. Note that we consider subtrees to be unordered and therefore instances of A(B)(C) and A(C)(B) are indexed under the same key. This can lead to incorrect results when joining results of a decomposed query in the subtree interval coding. For instance, a query such as A(C(D)(E))(B), can be evaluated using a join on the $C$ node of the results for A(C)(B) and those of C(D)(E). This join can be correctly evaluated, by selecting only the instance of A(C)(B), by consulting with the *order* values. Not filtering instances of A(B)(C), results in incorrect matches having the the structure A(B(D)(E))(C).

The structure of a posting describing a subtree of size $m$ is therefore as follows

$$\{tid, m, <l_1, r_1, v_1, o_1>, \ldots, <l_m, r_m, v_m, o_m>\}$$

where *tid* is an identifier of the tree that contain the subtree and $<l_i, r_i, v_i, o_i>$ values are the left, right, level and order numbers for the $i$th node, respectively. Remember that node orders for subtrees are according to a pre-order traversal.

### 4.4.3 Root-Split Interval Coding

The idea behind root-split (interval) coding is to avoid storing unnecessary structural information and to represent each subtree as concisely as possible. Root-split coding stores for each subtree only the tree identifier and *(pre, post, level)* values of its root. Compared to subtree interval coding, root-split coding reduces the size of each posting by a factor of at least $m$, where $m$ is the size of the subtree being indexed. Note that since the structural information of individual nodes are not stored in a root-split coding, the queries cannot be arbitrarily decomposed and joined. In the following, we define the types of covers required while evaluating queries over SI with root-split coding.

DEFINITION 8. *Given a query $Q$ and a cover $C$ of $Q$ such that $C = \{c_1, \ldots, c_k\}$, then $C$ is a root-split cover of $Q$ if and only if either $C = \{Q\}$ or for every subtree $c_i$, there exists a subtree $c_j$, $1 \leq i, j \leq k$, such that one of the following holds: (1) both $c_i$ and $c_j$ are rooted at the same node in $Q$, (2) $c_i$ is rooted at the parent of $c_j$ in $Q$, or (3) $c_j$ is rooted at the parent of $c_i$ in $Q$.*

Intuitively, a root-split cover is a cover which can be evaluated only by performing joins over the roots of its subtrees. Such a cover would be useful for our root-split coding as we only store structural information over roots of index keys. Every query $Q$ has at least one valid root-split cover, which is the set containing individual nodes of $Q$. In the next section we discuss how to construct (root-split) covers that lead to more efficient query execution times.

## 5. QUERY DECOMPOSITION

In this section, we study the theoretical properties of the root-split coding and compare it in terms of applicability and optimality with the subtree interval and filter-based codings. As will be discussed in this section, root-split coding reduces the size of posting lists, by reducing the number of postings and the size of each posting. As a result, the size of SI with root-split coding is smaller than its corresponding SI using subtree interval coding, by a large factor.

### 5.1 Monotonicity of Posting List Sizes

In the context of relational query optimization, intersection of the posting lists of subtrees indexed in SI, maps to select-project-join queries, with selections using index scan and joins using merge joins over sorted data streams (posting lists). In such a context a query optimizer over a subtree index, often generates query execution plans in the form of left-deep (or right-deep) trees resulting in a linear order of joins. Given a query $Q$, an efficient query plan can be obtained by (1) picking a "good" cover of $Q$ whose subtrees serve as data streams over leaves of the query plan, and (2) searching the space of available plans for the selected cover and finding an efficient or optimal query execution plan. The second step is the task of a query optimizer and we do not study it in this paper. However, in this section we study what properties of a cover make it more amenable for query optimization.

LEMMA 1. *For any two index keys $s_1$ and $s_2$ over a given SI, where $s_1 \precsim s_2$, we have*

  (i) *The posting list of $s_2$ is always a subset of the posting list of $s_1$ for filter-based coding.*

  (ii) *The posting list of $s_2$ is a subset of the posting list of $s_1$ for root-split coding if and only if $s_1$ and $s_2$ share the same root.*

  (iii) *The posting list of $s_2$ is not guaranteed to be a subset of the posting list of $s_1$ for subtree interval coding.*

PROOF. The proof is based on the structure of the three proposed codings.

  (i) In filter-based coding only the *tid* values are stored. If there exists a tree $t_k$ in the posting list of $s_2$ (i.e. $s_2 \precsim t_k$), since $s_1 \precsim s_2$, then we have $s_1 \precsim t_k$. The subtree relationship, $\precsim$, similar to subset relationship is transitive.

  (ii) For root-split coding, when $s_1 \precsim s_2$ and $s_1$ and $s_2$ share the same root, if there exists a tuple $T = \{t_i, < l_k, r_k, v_k >\}$ in the posting list of $s_2$, then it will be encoded using the same interval codings in the posting list of $s_1$. Therefore, $s_2$'s posting list is a subset of $s_1$'s.

  (iii) Proof for subtree interval coding is by a counter example. Assume we are given a SI with $mss = 2$ which has only the following tree indexed NP(NN)(NN)(NN). Apparently, NP $\precsim$ NP(NN), however, there are three entries in the posting list of NP(NN), while there is only one entry in the posting list of NP, which proves that the posting list of subtrees is not guaranteed to be supersets.

□

LEMMA 2. *For any two index keys $s_1$ and $s_2$ of a SI with root-split coding, where $s_1 \precsim s_2$ and $s_1$'s root has a different label from $s_2$'s root, then for each posting in the posting list of $s_1$ there is at most one posting in the posting list of $s_2$ associated with it.*

PROOF. Given the conditions of this lemma, $s_1$ must be a descendant of $s_2$'s root. Since ancestor-descendant relationship is a one to many relationship, there must be only one posting in the posting list of $s_2$ for any number of its descendants, hence the lemma is proved. □

The direct conclusion from Lemmata 1 and 2 is that the size of the posting lists monotonically decreases for subtrees of larger sizes in filter-based and root-split codings, while we do not have such a guarantee for the subtree interval coding. As a result, for a particular join, picking larger subtrees from the first two encodings guarantees a smaller join cost. In the next section we use this property to define max-covers and the notion of join optimality.

### 5.2 Join Optimality

As discussed earlier, root-split coding constrains query decomposition to covers in which subtrees can be joined over their roots only. In this Section, we investigate the ramifications of such a constraint on the size of the root-split covers. We study the number of joins required to evaluate a cover as a measure of its efficiency. Moreover, we study the problem of join optimality for root-split and non root-split covers.

#### 5.2.1 Max Covers

Given a query $Q$, a valid root-split cover $C$ over $Q$ might have subtrees ranging in size from 1 to $mss$. One interesting problem is to investigate if there exists an algorithm that can always generate a root-split cover, where the size of every subtree is equal to *mss*. We call such a cover a *max-cover*. According to the discussion in the previous section, such a cover would achieve an efficient query evaluation plan. Among all max-covers of $Q$, only a few are root-split, and among such max-covers, those with the smallest size, in terms of the number of subtrees, are desirable as they lead to our definition of a join-optimal cover.

DEFINITION 9. *For a given query $Q$, a join-optimal cover of $Q$, is a max-cover over $Q$ that has the smallest size in terms of the number of subtrees among all covers of $Q$.*

Note that for a cover $C$ of $Q$ and a SI with either of filter based or root-split codings, there always exists a max-cover which has size smaller than or equal to $|C|$. As a result, we do not need to check join optimality for covers that are not max-covers. However, according to Lemma 1, it is not always desirable to select covers over a subtree interval coding from max-covers. As that lemma shows, larger subtrees in the covers do not necessarily lead to smaller posting lists, when subtree interval coding is used. However, an experiment we conducted showed that max-covers are often good heuristics for evaluating queries over subtree interval coding scheme. On a dataset containing more than $112,000$ index keys of sizes $1, \ldots, 5$, in only $0.0005\%$ of cases there exist index keys $k_i$ and $k_j$, such that $k_i \precsim k_j$ and $k_j$ has a larger posting list than $k_i$. Therefore, in order to significantly reduce the search space for good covers, we only consider max-covers (hereafter covers) over all proposed coding schemes.

### 5.2.2 Join Optimal Covers

In this section, we study the problem of finding join optimal covers.

DEFINITION 10. *Given a query $Q$, and a cover $C$ over it, we say that $C$ has deep branching anomaly, if there exist subtrees $s_i$ and $s_j$ in $C$ such that (1) $s_i$ and $s_j$ share at least one node of $Q$, say $v \in V(Q)$, such that $v$ is not root of $s_i$, and $v$ is not root of $s_j$, and (2) $v$ has at least two children $u$ and $u'$, such that $u \in V(s_i)$, $u \notin V(s_j)$ and $u' \in V(s_j)$ and $u' \notin V(s_i)$.*

Deep branching anomaly, as defined in Definition 10, describes a situation where two subtrees in a given cover cannot represent the structure of part of the query they cover, uniquely. Deep branching anomaly can result in extraneous matches for root-split codings. As a result of a deep branching anomaly, extra subtrees might be required to be added to the root-split covers to fix this anomaly.

For non root-split codings, deep branching anomaly can be dealt with, efficiently. Such an anomaly does not introduce any issues for query evaluation under filter-based coding as the final set of matches is computed by scanning over the candidate matches. For subtree interval coding this anomaly can be dealt with by joining on the deepest shared branching node of the two subtrees. In the example that follows, we demonstrate how it is possible to handle deep branching for subtree interval coding.

EXAMPLE 1. *Consider the query in Figure 5.(a) and let $mss = 4$. A join-optimal root-split cover of the query is $C_1$={A(B(C(D))), B(C(E)(F))}. Figure 5.(b) shows multiple tree structures that match the given root-split cover. The result set obtained by an anomalous join over roots of the subtrees in $C_1$, i.e.* A *and* B *nodes, thus can result in extraneous matches for root-split coding. By sacrificing join optimality, we can obtain root-split covers that do not have the deep branching anomaly, as in the following cover $C_2$={A(B(C(D))), B(C(E)(F)), C(D)(E)(F)}. Over subtree interval coding, join optimality can still be achieved by performing the joins on the deepest branching node, i.e.* C.

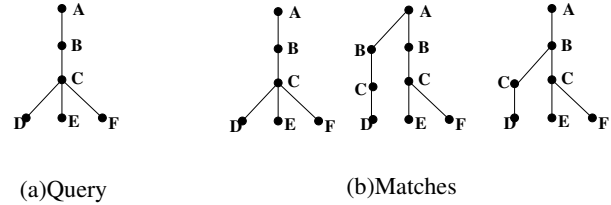

(a)Query          (b)Matches

**Figure 5: Example of a query having deep branching anomaly with $mss = 4$**

PROPOSITION 1. *The number of extra joins required for evaluating a root-split cover of a query $Q$ is at most $|Q| - \lceil \frac{|Q|}{mss} \rceil - mss + 1$, compared to a join optimal cover.*

PROOF. The worst case happens when the tree is structured as a unary branch of height $|Q|$. In this case, the number of subtrees for a root-split cover is given by $|Q| - mss + 1$, while the number of subtrees in a join optimal cover is given by $\lceil \frac{|Q|}{mss} \rceil$. The difference of these two terms gives our proposition bound. □

The above proposition provides an untight upper bound on the number of extra joins required to evaluate a root-split cover. In practice however, as we will show in Section 6.3, the actual number of extra joins is smaller than the above bound. In the rest of this section, we provide algorithms for computing covers for both root-split and non root-split codings.

Figure 6 shows the *optimalCover* algorithm that computes a join optimal cover for the input query $Q$. The algorithm starts with an empty cover $C$, and in each step either adds a subtree of size $mss$ or calls *optimalCover* on larger subtrees. Thus, at the base of the recursion, *optimalCover* handles only children of $Q$ having size less than or equal to $mss$. Any child of $Q$ with size equal to $mss$ is added immediately to the cover $C$. Children with sizes smaller than $mss$ are handled by calling *assign* until the total number of unassigned nodes in the subtree of $Q$ and including $Q$ is less than $mss$. At this point if $Q$ is not the root of the original query, $Q$ and its unassigned nodes can be part of a subtree originating from parent of $Q$, and thus the *optimalCover* returns. Otherwise, if $Q$ is the root of the original pattern, all that is left to do is to cover the last set of unassigned nodes, whose number is less than $mss$. This is achieved by one last call to *assign* in lines $9 - 10$. The algorithm *assign* is also presented in the same figure. Intuitively, a call of $assign(t)$ computes a subtree of size $mss$, rooted at $t$, which has the most possible set of unassigned nodes. It starts by picking larger unassigned children of $t$ and once it runs out of unassigned nodes, adds assigned nodes until the size of subtree is $mss$.

**optimalCover(Q)**
1  $C \leftarrow \emptyset$
2  **for** $c \in Q.children$
3    **if** $|c| = mss$
4      $C \leftarrow C \cup c$, $|Q| = |Q| - |c|$, $c.assigned = true$
5    **else if** $|c| > mss$
6      $C \leftarrow C \cup $ **optimalCover(c)**
7  **while** $|Q| \geq mss$
8    $C \leftarrow C \cup $ **assign(Q)**
9  **if** $|Q| > 0$ and $isRoot(Q)$
10    $C \leftarrow C \cup $ **assign(Q)**
11 **return** $C$

**assign(Q)**
1  $cnt = 1$, $t.root = Q.root$, $Q.assigned = true$
2  sort $Q.children$ on size, descending
3  **for** $c \in Q.children$
4    **if** $c.assigned = false$
5      **if** $(mss - cnt - |c|) \geq 0$
6        $c.assigned = true$, $t.children \leftarrow t.children \cup c$
7        $|Q| = |Q| - |c|$, $cnt = cnt + |c|$
8    **if** $cnt = mss$ **then return** $t$
9  **if** $cnt < mss$
10   **for** $c \in (Q.children - t.children)$
11     **if** $(mss - cnt - |c|) \geq 0$ **then**
12       $t.children \leftarrow t.children \cup c$, $cnt = cnt + |c|$
13     **else**
14       add **subtrees(c, mss − cnt)[0]** to $t.children$
15 **return** $t$

**Figure 6: Algorithm that computes a join optimal cover of size $mss$**

EXAMPLE 2. *Consider the tree shown in Figure 1.(a) and suppose we run the algorithm optimalCover on this tree with $mss = 3$. The first child of S is NP(NNS(agouti)) of size 3 and this child is added to C immediately. The second child of S, VP, is of size 7, so optimalCover(VP) is called, which in turn calls optimalCover on NP of size 4. Since DT(a) and NN both have size less than mss, assign(NP) is called; the call returns NP(DT(a)) which is added to C and sets $|NP| = 2$. Since NP is not a root (line 9 of optimalCover), C is returned to the caller. The next steps of the algorithm will add VP(VBZ(is)), VP(NP(NN)) and S(NP(NNS)) to the cover. Note that a join of subtrees VP(NP(NN)) and NP(DT(a)) must be in the form of an equality join on node NP, to avoid erroneous results due to deep branching anomaly.*

LEMMA 3. *Given a parameter $mss \leq 6$ and a tree $t$, where $|t| > mss$ and all children of $t$ have size less than $mss$, repeated calls of assign over $r(t)$ partitions $t$ into a join optimal cover.*

PROOF. Since children of $t$ all have size less than $mss$, any subtree that covers them has to be rooted at $r(t)$. Thus, the partitioning problem reduces to the integer bin packing problem, where the bin capacity is $mss - 1$ and children sizes are the volumes of the items to be stored. The objective is to minimize the number of bins (subtrees in our problem). Our *assign* algorithm sorts children in a non-increasing order of their sizes, which maps to the *fit first decreasing (FFD)* approximation algorithm for bin packing. FFD in general gives approximation ratio of $\frac{11}{9}OPT + 1$ [21] and is shown to be optimal for integer bin packing with bin sizes less than or equal to 6, which proves our lemma. □

The above lemma proves that for small values of $mss$, assign provides an optimal partitioning and for general $mss$, it achieves a good approximation ratio. As discussed earlier, the number of extracted subtrees could grow dramatically as $mss$ increases and therefore, in practice we will not be dealing with $mss$ values larger than 6. In our experiments, we limited $mss$ to be at most 5.

THEOREM 1. *Given that the size parameter $mss \leq 6$, then optimalCover computes a join optimal cover.*

PROOF. We assume that $|Q| \geq mss$, otherwise, $Q$ can be covered using a single subtree, which is obviously join optimal. *optimalCover* starts from the root of $Q$. For each child $c$ of $Q$, we have one of the following three cases, (1) $|c| < mss$, (2) $|c| = mss$, and (3) $|c| > mss$. Case (1) is handled by *assign* algorithm which we showed join optimality in Lemma 3. Case (2) is directly assigned into an individual subtree partition at line 3 of the *optimalCover*. Finally, case (3) is handled by recursive calls of *optimalCover* until either of cases (1) or (2) occur. Over internal nodes with condition of case (1), as soon as enough of their children are assigned and their remaining size reduces to less than $mss$, *optimalCover* returns and leaves their handling to the ancestor which satisfies case (1). As a result *optimalCover* achieves a globally join optimal cover over $Q$. □

Through some modifications of the *optimalCover* algorithm, we can develop an algorithm that obtains the smallest root-split cover in terms of size. This algorithm, referred to as *minRC*, is presented in Figure 7. This new algorithm takes a bottom-up approach and descends into subtrees of smaller sizes until children have size less than or equal to $mss$. Then, it covers the given subtree entirely, before moving up to higher levels. This guarantees that every child of a given node $v$ is covered, before $v$ is covered and as a result deep branching anomaly is avoided.

**minRC(Q)**
1  $C \leftarrow \emptyset$
2  **for** $c \in Q.children$
3    **if** $|c| = mss$
4      $C \leftarrow C \cup c$, $|Q| = |Q| - |c|$, $c.assigned = true$
5    **else if** $|c| > mss$
6      $C \leftarrow C \cup $ **minRC(c)**
7  **while** $|Q| \geq 0$
8    $C \leftarrow C \cup $ **assign(Q)**
9  **return** $C$

**Figure 7: Algorithm that computes the best root-split cover of size $mss$**

EXAMPLE 3. *The minRC algorithm returns the following cover over the query in Figure 1.(a). $C = \{$NP(NNS(agouti)), NP(DT(a)), NP(DT)(NN), VP(VBZ(is)), S(NP(NNS))$\}$. The subtree ordering shown is the same as the order by which minRC adds subtrees to C. C is join optimal, and it has the same number of subtrees as an optimal cover, given in Example 2.*

THEOREM 2. *Given a parameter $mss \leq 6$, minRC returns the smallest root-split cover possible.*

PROOF. *minRC* handles internal nodes that fall into case (1) of the proof in Theorem 1 different from *optimalCover*. To avoid deep branching anomaly, it requires that each internal node is assigned to a subtree, before any of its ancestors are assigned. As a result, there are cases where *minRC* does not achieve optimality. However, since all root-split covers have to handle deep branching anomaly, *minRC* achieves the smallest cover possible among them, by repeatedly calling *assign* on non-assigned subtrees, which was shown to be optimal. □

## 6. EXPERIMENTAL EVALUATION

### 6.1 Experimental Setup

For our experiments in this section, we parsed a collection of sentences from the AQUAINT corpus of English News Text [2], using Stanford Parser [9], and used this dataset or a part of it in our experiments. This dataset would be a good representative for any parsed English dataset as the same grammar and similar rules are applied when parsing, and the set of tags and their corresponding disctributions are very similar. We further processed the collection of obtained parse trees and assigned ids and structural tags to their nodes. With this tagging, each node is described as a tuple consisting of *treeId, nodeId, parentId, pre, post, level* and *label*. The treeId value points to the corresponding tree that contains the node. The *nodeId* is a numeric value that uniquely identifies each node within a tree, and *parentId* is the *nodeID* of the parent node.

We constructed two sets of queries over syntactically annotated text for our experiments. The first set, *WH query-set* was created by a third person who was asked to select 48 questions from AOL query log [12], 12 questions from each of *what, which, where* and *who* questions. She was then asked to rewrite the questions in the form of matching sentences. For instance a question such as *who is the mayor of New York city?* could be converted to *mayor of New York city is %match%*. Finally, we parsed these sentences using Stanford parser and removed for each sentence the leaves that contain terms from the sentence, leaving only the sentence structure.

Our second query set, *FB query-set*, was constructed by extracting subtrees from a set of parsed sentences which were not included in our indexes. The extracted subtrees were selected according to the frequencies of their nodes. To account for differences in the selectivities of queries that are posed to our indexes, we constructed 7 classes of queries, consisting of nodes with high (H), medium (M) and low (L) frequency labels, together with their combinations; i.e. HM, HL, ML, HML. For each class, we construct 10 subtrees of size 1 to 10.

Our subtree index was implemented as a native disk-based B+Tree index. We did not implement a caching system over the B+Tree and relied on the page buffering of the operating system for any savings in the number of disk page accesses. We also flattened and sequentially stored parse trees in a separate file, which we call the *data file*. All our experiments were run on a 64-bit machine with 64 GB of physical memory and a 4x quad-core processor. The system page size was 4096 bytes. The reported index sizes will be different on 32-bit addressing systems or with different page sizes. More details of the setup is presented in [7].

### 6.2 Index Construction

In this section we study the characteristics of the indexes built over syntactically annotated trees experimentally. We investigate how the size of the index is affected by the choice of the coding scheme and size of the input data. We also study the index construction time for different coding schemes and input sizes.

#### 6.2.1 Index Size

Figure 8 shows the subtree index size for the three proposed coding schemes and varying input sizes, with the input size marked on top of each sub-figure. Furthermore, in each sub-figure, we vary the maximum subtree size, $mss$, from 1 to 5, as shown on the X axis.

As the figure shows, the size of the index is smallest for filter-based coding, and largest for the subtree interval coding in all cases. One interesting pattern in the results for sizes of the index is that as $mss$ increases, the gap between the sizes of root-split and subtree interval codings grows. The reason is that for larger subtrees, subtree interval coding uses larger postings, because it has to store the structural information for individual nodes. However, the posting size in root-split coding has constant size, and the index size increases only due to more keys being indexed.

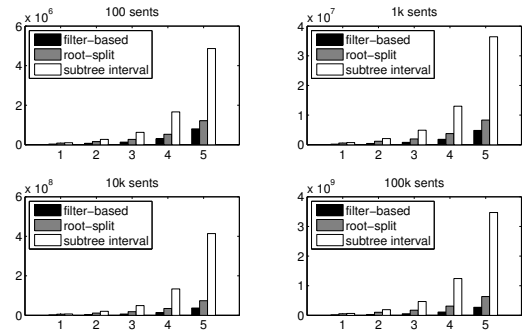

**Figure 8: SI size (bytes) for filter-based, root-split and subtree interval codings, with $mss = 1, \ldots, 5$.**

Table 1 shows the ratio of the index size when $mss$ is 5 to the the index size when $mss$ is 1, for all three coding schemes and four dataset sizes. As the table depicts, root-split coding shows the smallest increase in the size of the index among all coding schemes.

**Table 1: Ratio of the subtree index size when $mss$ is 5 to the index size when $mss$ is 1**

|        | Filter-based | Root-split | Subtree Interval |
|--------|--------------|------------|------------------|
| 100    | 22           | **15**     | 48               |
| $1k$   | 24           | **14**     | 50               |
| $10k$  | 23           | **13**     | 59               |
| $100k$ | 21           | **12**     | 54               |

The size reduction for root-split coding is due to (1) reducing the size of each posting as only structural information of roots are stored, and (2) reducing the number of postings as multiple subtrees which have the same key and the same root structural information will be represented with only one posting in root-split coding, while every single subtree requires a distinct posting using the subtree interval coding. Figure 9 depicts the number of postings for our three coding schemes, varying the dataset size and $mss$. As this figure shows, for $mss = 1$ the number of postings of root-split and subtree interval codings are equal and as $mss$ increases the gap between the number of postings for these coding schemes widens. Filter-based coding has the smallest number of postings as it only stores unique $treeId$s, and no structural information.

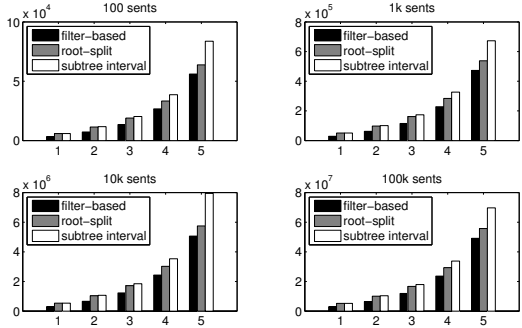

Figure 9: **Total number of postings for filter-based, root-split and subtree interval codings, with varying input sizes and $mss$ values.**

Finally, to have an idea of the space overhead of the index, the size of a B+tree constructed over subtree inverted lists is comparable to the size of the data file for $mss = 1$. For larger values of $mss$, the gap between the data file size and subtree index size grows. For $mss = 5$ and subtree interval coding, the size of data file is two orders of magnitude smaller than the subtree index size.

### 6.2.2 Index Construction Time

Figure 10 shows the construction time of the subtree index for different datasets, coding schemes and $mss$ values. As shown, the construction time is smallest for filter-based coding and largest for subtree interval coding. Root-split has a construction time that is slightly larger than filter-based coding. As $mss$ increases the difference in the construction time between subtree interval coding and the other two codings grows. This is mostly because the size of the index for subtree interval coding is larger and as a result more data has to be written over disk.

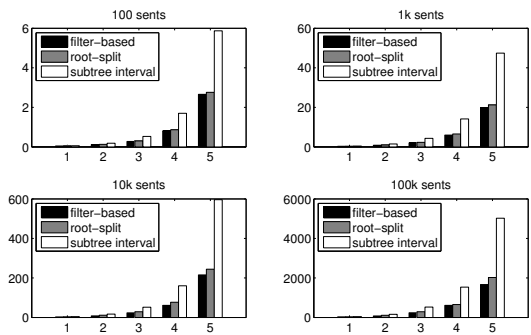

Figure 10: **Index construction time (seconds) for filter-based, root-split and subtree interval codings, with $mss = 1, \ldots, 5$ and varying input sizes.**

## 6.3 Querying Performance

In this section, we experimentally evaluate the performance of querying of our subtree index under different settings. In particular, we investigate the runtime of queries in terms of their number of matches for the filter-based, root-split and subtree interval codings as $mss$ values and query sizes vary. We also compare the performance of our subtree index using root-split coding with ATreeGrep [16] and our adaptation of TreePi [23] over tree structured data, which we call frequency-based approach. We also present some scalability results for the three coding schemes using data sizes of one thousand to one million sentences. Finally, we study the performance of our decomposition algorithms by comparing the number of joins that are required under each query decomposition policy.

### 6.3.1 Response Time of Queries

To obtain the query response time over our subtree index, we used all the 48 WH queries and 70 FB queries, and tried each query 5 times and took the average running time per query. We grouped the queries according to their total number of matches into the following bins: (1) less than 10, (2) between 10 and 100, (3) between 100 and $1k$, (4) between $1k$ and $10k$ and (5) larger than $10k$ matches. Figure 11 shows the average run-time of queries with varying number of matches on the horizontal axis, over $100k$ sentences.

As Figure 11 shows, the running time of the queries decreases for all coding schemes as $mss$ increases. This reduction is smallest for queries with large number of matches using filter-based coding, as the time of the *filtering* phase becomes a dominating factor, when there are many matches. As shown in the figure, Root-split coding performs better than subtree interval coding in all cases. Filter-based coding performs better than root-split coding for $mss = 1$ and less than 10 matches on average. However, for larger values of $mss$, which are mainly interesting for a subtree index, root-split coding performs better than the other two coding schemes.

Unlike the filter-based coding, both the root-split and subtree interval codings display a reduction in their average query response times for larger number of matches, for the following reasons. (1) The intermediate result size of a query with a small number of matches could be large and this would affect the runtime of queries under root-split and subtree interval codings, but not under filter-based coding. (2) As is expected, larger queries have on average smaller number of matches; however, these queries require a larger number of joins and take longer for these two coding schemes. This pattern can also be seen in Figure 12 where the running time is depicted as the query size varies.

Figure 12 displays the runtime of queries in terms of the query size using the same settings as in Figure 11. In this figure, we only included queries which have 100 and more matches. As this figure shows, root-split and subtree interval codings show an increasing trend with respect to the size of queries. Filter-based coding displays a somewhat random behavior with respect to the query size as its performance is mostly determined by the number of matches and how well the cover subtrees can perform filtering. According to this figure, as $mss$ increases, root-split and subtree interval codings perform better on larger queries as they require smaller number of joins to compute the result set of queries.

### 6.3.2 Comparison with Other Systems

Table 2 displays the results of comparing our SI with $mss = 3$ using root-split coding with ATreeGrep [16] and a frequency-based approach that is an adaptation of TreePi [23]

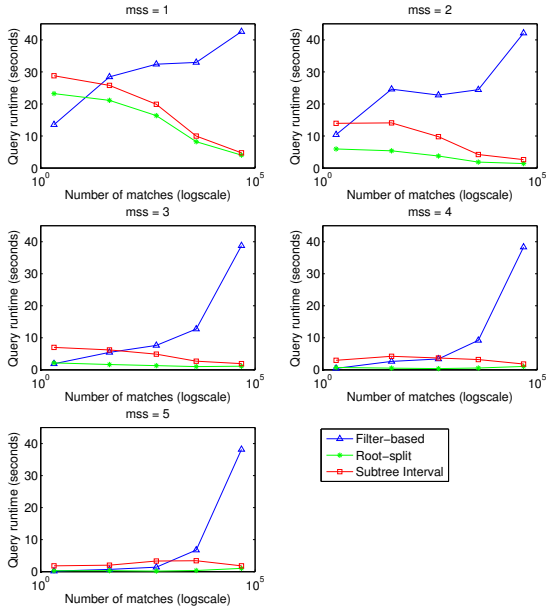

**Figure 11: Average runtime of queries in terms of number of matches for filter-based, root-split and subtree interval codings and $mss$ values of 1 to 5**

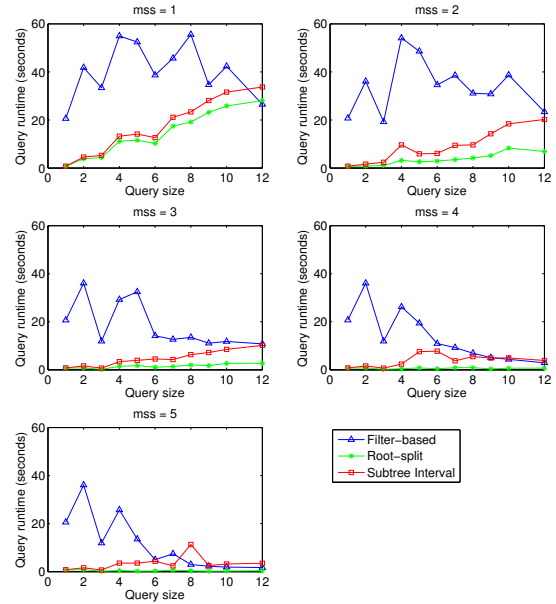

**Figure 12: Average runtime of queries in terms of the size of queries for filter-based, root-split and subtree interval codings and $mss$ values of 1 to 5**

for indexing parse trees. Similar to TreePi, the frequency-based approach stores in the index all single nodes and a percentage of larger highest frequency subtrees. This percentage is denoted in brackets in the last three columns of Table 2. The frequency-based approach also uses the same query decomposition algorithm as TreePi, except over tree structured queries.

The results in Table 2 are obtained over the queries in our *FB query-set* and are grouped by the frequency classes. Since ATreeGrep does not support all the queries, the results are averaged over as many queries as there were results for. As this table depicts, SI with root-split coding outperforms by at least one order of magnitude over all frequency classes.

**Table 2: Average runtime of FB query classes using Subtree index with root-split coding ($mss = 3$), ATreeGrep and Frequency-based approaches with varying frequency cutoff thresholds.**

|     | **RS** | **ATG** | **FB**(0.1%) | **FB**(1%) | **FB**(10%) |
|-----|--------|---------|--------------|------------|-------------|
| L   | **0.09** | 1.9   | 3.05         | 3.03       | 3.04        |
| M   | **0.01** | 10.06 | 12.32        | 0.8        | 0.35        |
| ML  | **0.25** | 2.13  | 10.3         | 9.62       | 9.25        |
| H   | **1.73** | 22.4  | 39.21        | 34.51      | 34.53       |
| HL  | **1.57** | 32.97 | 34.58        | 34.61      | 34.6        |
| HM  | **1.76** | 37.08 | 35.54        | 31.40      | 31.57       |
| HML | **1.76** | 86.02 | 49.03        | 42.97      | 43.13       |

### 6.3.3 Scalability Results

Figure 13 presents the average runtime of our queries over four subsets of our parsed collection. $1k$, $10k$, $100k$ and $1m$ sentences. We used $mss = 3$ for the results reported in this figure, but the result for other values of $mss$ were similar. The reported runtimes are the average query response times over all queries in *FB query-set* and using our three coding schemes. The results in this figure show that all three coding schemes display a similar pattern as the dataset size increases, i.e. the running time grows approximately linearly with the number of sentences indexed.

Figure 13 also shows that the root-split coding scales up better with the dataset size. Averaged over all queries and in the range of $1k$ to $1m$ sentences of our parsed collection, the query runtime increases for filter-based, subtree interval and root-split codings by a factor of 1025, 752 and 529, respectively.

### 6.3.4 Query Decomposition Algorithms Results

Table 3 displays the average number of joins per query over groups of *Who*, *Which*, *Where* and *What* queries for $mss$ values of 2 to 5. As Table 3 shows, *optimalCover* achieves a fewer number of joins for all groups of queries and $mss$ values[2]. Despite a fewer number of joins obtained for filter-based and subtree interval codings, root-split still manages to have a smaller query response time, by minimizing the I/O cost and avoiding to perform filtering.

## 7. CONCLUSIONS AND FUTURE DIRECTIONS

We proposed a novel subtree index in this paper. Subtree index improves the query response time of node and edge approaches by reducing the posting lists and the number of joins required for evaluating tree structured queries. Compared to path approaches, it preserves the structure of trees within index keys, and can perform exact matchings without a need for post validations. Our root-split coding provides a concise encoding of the structural information of subtrees and reduces the index size, index construction time and query response time compared to one or both of our baseline methods. It also improves the querying performance by at least an order of magnitude compared to previous approaches.

---
[2]In the case where $mss = 1$, root-split and subtree interval will have equal number of joins, which is equal to $|Q| - 1$

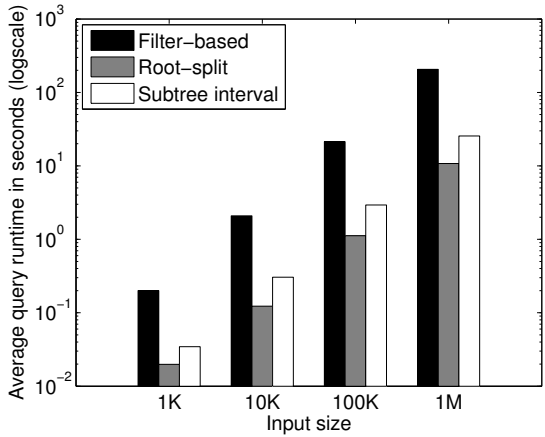

**Figure 13: Average runtime of queries ($mss = 3$) over datasets of $1k$, $10k$, $100k$ and $1m$ sentences and using different coding schemes.**

**Table 3: Average number of joins required over queries in the WH query set. r=root-split, s=subtree interval.**

| Query set | $mss = 2$ | | $mss = 3$ | | $mss = 4$ | | $mss = 5$ | |
|---|---|---|---|---|---|---|---|---|
| | r | s | r | s | r | s | r | s |
| Who | 5.91 | 5.41 | 4.75 | 3.33 | 3 | 2.16 | 2.41 | 1.66 |
| Which | 6.83 | 6.25 | 5.41 | 4.25 | 4.25 | 3 | 3.25 | 2.25 |
| Where | 4.91 | 4.75 | 4.41 | 3.33 | 2.66 | 2.08 | 2.25 | 1.58 |
| What | 5.58 | 5.33 | 4.58 | 3.33 | 2.91 | 2.25 | 2.25 | 1.58 |

As future improvements we propose adapting more efficient structural join approaches such as TwigStack [5] over our subtree index. Moreover, it would also be interesting to further study the problem query optimization over SI, by considering building data structures that store statistics about subtrees such as their selectivities.

## 8. ACKNOWLEDGMENTS

This research was supported by the Natural Sciences and Engineering Research Council and the BIN network.